\documentclass[useAMS,usenatbib]{mn2e}
\usepackage{graphicx}
\usepackage{aas_macros}
\title{Differential chemical abundance analysis of a 47 Tuc AGB star with
respect to Arcturus}
\author[C.C. Worley et al.]
       {C.C. Worley$^{1}$\thanks{E-mail: clare.worley@pg.canterbury.ac.nz (CCW)}, P.L. Cottrell$^{1}$, K.C. Freeman$^{2}$ and E.C. Wylie-de Boer$^{2}$\\
$^{1}$The Beatrice Tinsley Institute, Dept. of Physics \& Astronomy,
University of
Canterbury, Private Bag 4800, Christchurch, New Zealand\\
$^{2}$Research School of Astronomy \& Astrophysics, Australian
National University, Cotter Rd, ACT, Australia}

\begin{document}

\date{Accepted for publication in MNRAS}

\pagerange{\pageref{firstpage}--\pageref{lastpage}} \pubyear{2009}

\maketitle

\label{firstpage}

\begin{abstract}
This study resolves a discrepancy in the abundance of Zr in the 47
Tucan\ae\ asymptotic giant branch star Lee~2525. This star was
observed using the echelle spectrograph on the 2.3~m telescope at
Siding Spring Observatory. The analysis was undertaken by
calibrating Lee~2525 with respect to the standard giant star
Arcturus. This work emphasises the importance of using a standard
star with stellar parameters comparable to the star under analysis
rather than a calibration with respect to the Sun \citep{Koch2008}.
Systematic errors in the analysis process are then minimised due to
the similarity in atmospheric structure between the standard and
programme stars. The abundances derived for Lee 2525 were found to
be in general agreement with the \cite{Brown1992} values except for
Zr. In this study Zr has a similar enhancement
([Zr/Fe]~$=+0.51$~dex) to another light {\it s}-process element, Y
([Y/Fe]~$=+0.53$~dex), which reflects current theory regarding the
enrichment of {\it s}-process elements by nuclear processes within
AGB stars \citep{Busso2001}. This is contrary to the results of
\cite{Brown1992} where Zr was under-abundant ([Zr/Fe]~$=-0.51$~dex)
and Y was over-abundant ([Y/Fe]~$=+0.50$~dex) with respect to Fe.
\end{abstract}


\section{Introduction}\label{sec:Intro_BW}
The globular cluster 47~Tucan\ae\ (47~Tuc) has proven to be a rich
source of study with regards to the structure and evolution of stars
and stellar systems. Chemical abundance studies have indicated the
presence of light element abundance anomalies between the stars at
all stages of stellar evolution. Recent advances in telescope and
instrument sensitivity are providing observation of larger samples
of stars within clusters at higher resolution enabling a more
detailed investigation of the exact nature of the abundance
anomalies. A significant advance has been the greater number of
abundances derived for the elements heavier than iron. In particular
the light and heavy {\it s}-process elements which provide
signatures of key stages in stellar evolution.

A well studied phenomenon in 47~Tuc is the CN weak, CN strong
bimodality which is seen at all stages of stellar evolution in
47~Tuc \citep{Cannon1998}. Internal mixing cannot solely explain
this anomaly and some primordial or pollution mechanism is required
to account for it being observed on the main sequence and giant
branches (\citealt{Cannon1998}; \citealt{Briley2004}). Also observed
in 47~Tuc is a correlation of Na to CN strength
\citep{Cottrell1981}. However there has been no observational
evidence of a correlation of either Mg or Al with CN or Na as can be
observed in more metal poor clusters though Na has been observed to
anti-correlate with O in 47 Tuc \citep{Carretta2004}. As these
anomalies have been observed in giants and dwarfs it is theorised
they are most likely due to primordial scenarios rather than solely
due to internal mixing in globular cluster stars. For a complete
summary of these anomalies in 47 Tuc and other globular clusters
refer to the recent review paper, \cite{Gratton2004}.

Recent stellar studies have investigated the abundances of elements
heavier than iron and have shown an enhancement in {\it s}-process
elements in 47~Tuc. \cite{Brown1992} analysed four giant stars in
47~Tuc for their light and heavy element abundances, determining for
the {\it s}-process elements an average enhancement in Y
([Y/Fe]~$=+0.48\pm0.11$~dex) but a depletion in Zr
([Zr/Fe]~$=-0.22\pm0.05$~dex). \cite{Alves-Brito2005} analysed a
sample of five 47~Tuc giant stars and found Zr to also be depleted
([Zr/Fe]~$=-0.17\pm0.12$~dex) though did not obtain abundances for
Y. However \cite{Wylie2006} found an overall enhancement in {\it
s}-process elements, including Zr ([Zr/Fe]~$=+0.65\pm0.16$~dex) and
Y ([Y/Fe]~$=+0.64\pm0.20$~dex), in seven giant branch stars in 47
Tuc. In a study of 3 turn-off and 8 subgiants in 47~Tuc
\cite{James2004b} found enhancements in Sr and Ba, but Y was
depleted for the subgiants ([Y/Fe]~$=-0.11\pm0.10$~dex) and slightly
enhanced for the turnoff stars ([Y/Fe]~$=+0.06\pm0.010$~dex). While
the results are consistent within each study the variation between
studies are contradictory to the expected homogeneous distribution
of heavy elements in globular cluster stars.

The primary source for {\it s}-process element production are
nuclear reactions that occur during the asymptotic giant branch
(AGB) stage of stellar evolution \citep{Busso2001}. AGB stars of
mass greater than 1.5~M$_{\odot}$ undergo third dredge up (TDU)
during which seed nuclei (Fe) are exposed to neutron fluxes. This
builds up heavier and heavier nuclei resulting in the observed
enhancements in {\it s}-process element abundances in thermally
pulsing AGB stars. Due to small reaction cross-sections, elements
are first built up in the light {\it s}-process peak ({\it ls} =
$\langle$Sr, Y, Zr$\rangle$) then in the heavy {\it s}-process peak
({\it hs} = $\langle$Ba, La, Nd$\rangle$) \citep{Busso2001}. The
ratios of [{\it ls}/Fe], [{\it hs}/Fe] and [{\it hs/ls}] are used as
indicators of the degree of {\it s}-process enhancement and neutron
flux. A depletion in Zr coinciding with an enrichment in Y, as
determined in \cite{Brown1992}, is not a likely result in a scenario
of enrichment due to AGB stars. There are other potential sources
for the {\it s}-process elements. The AGB source is referred to as
the `main' {\it s}-process. The `weak' {\it s}-process occurs during
He-core burning in massive stars and can enhance light {\it
s}-process elements such as Sr and Y \citep{Arlandini1999}. The {\it
r}-process, rapid neutron exposures typically in supernov\ae, can
also contribute to the abundances of {\it s}-process elements. The
contributions from these different sources can be disentangled by
comparing heavy elements whose abundances are mainly {\it s}-process
(such as Y and Zr) to those who are mainly {\it r}-process, such as
Eu, to those who have contributions in different ratios from these
sources \citep{Travaglio2004}.

While there are these different potential sources for each of the
{\it s}-process elements, including contributions from the {\it
r}-process, for the most part it is argued that the abundance of a
heavy element is homogeneous within a cluster. The differing results
for Zr (and Y) need to be resolved in order to pursue the
connections between heavy and light element abundances. These
connections provide clues as to the nature of primordial and mixing
scenarios that could lead to chemical abundance distributions that
are seen in globular cluster stars.

\section{Observations and previous work}\label{sec:observations}
In November 2007 the 47~Tuc giant branch star Lee~2525 was observed
on the 2.3~m telescope at the Siding Spring Observatory by K.C.
Freeman and E.C. Wylie-de~Boer. It was observed using the
\'{e}chelle spectrograph, obtaining wavelengths from 4500~\AA\ to
6500~\AA. The spectrum had a resolution of $R\sim 20,000$ with a
signal to noise per resolution element of $\sim$35. The spectrum was
reduced in IRAF and the normalised spectrum underwent the same
equivalent width analysis as the calibration star Arcturus (see
Section~\ref{sec:ArcCalib}). Lee 2525 had been observed previously
in \cite{Brown1992}, along with four other 47~Tuc giant branch
stars. In \cite{Wylie2006} these stars were compared with seven more
giant stars in 47~Tuc.

\begin{figure}
%\begin{minipage}{180mm}
\begin{center}
\includegraphics[width = 90mm]{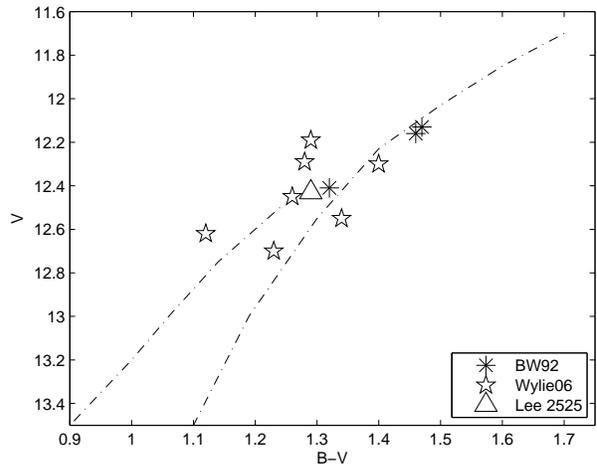}
\caption{Colour-magnitude diagram of 47~Tuc showing the placement of
the giant stars observed in \citet{Brown1992} and \citet{Wylie2006}.
The fiducial lines indicating the RGB and AGB branches of 47~Tuc are
from \citet{Hesser1987}.} \label{fig:BWWylieCMD}
\end{center}
%\end{minipage}
\end{figure}

Figure~\ref{fig:BWWylieCMD} shows the placement of the stars in both
studies on the colour-magnitude diagram of 47~Tuc. Lee~2525 is also
indicated on the diagram and all of the stars fall on the red giant
branch (RGB) or AGB as indicated by the fiducial lines
\citep{Hesser1987}.

Lee~2525 has been studied in a number of papers. As a bright giant
it was initially observed under this designation in \cite{Lee1977}.
Subsequently it was studied in \cite{Norris1979} where it was
determined to be a CN weak AGB star. It has also been part of a mass
loss rate study for giant stars in 47~Tuc where it was determined to
have no infrared excess \citep{Ramdani2001}. Lee~2525 was analysed
in \cite{Brown1992} as the CN weak star in a CN weak-CN strong pair.
Its counterpart was Lee~1513.

\cite{Worley2008} observed Lee~2525 as part of a study into the
feasibility of medium-resolution surveys of globular cluster stars
to determine their {\it s}-process elemental abundances using the
Robert Stobie Spectrograph (RSS) on the Southern African Large
Telescope (SALT). For Lee~2525 in particular there was an indication
of an enhancement in Zr, although this lay within the uncertainties
of the model abundance. The resolution of the spectra in this study
was determined to be too low (R$\sim$5,000) to determine absolute
{\it s}-process element abundances, although upper limits of
$0.5$~dex may be possible in future higher resolution observations
with RSS.

The average heavy elemental abundances for the stellar samples
reported in \cite{Brown1992}and \cite{Wylie2006} are listed in
Table~\ref{tab:Wylie_Browncomp}.

\begin{table}
\caption{Comparison of the mean heavy elemental abundances from
samples of AGB and RGB stars analysed in \citet{Brown1992} (BW92)
and \citet{Wylie2006} (W06).} \label{tab:Wylie_Browncomp}
%\begin{minipage}{80mm}
\centering
% Table generated by Excel2LaTeX from sheet 'BWWylie'
\begin{tabular}{ccc}

  & BW92 & Wylie06 \\

No. Stars & 4 & 7 \\

$\langle$[Fe/H]$\rangle$ & $-0.81$ & $-0.63$ \\

  &   &   \\

$\langle$[Y/Fe]$\rangle$ & $+0.48$ & $+0.64$ \\

$\langle$[Zr/Fe]$\rangle$ & $-0.22$ & $+0.65$ \\

$\langle$[La/Fe]$\rangle$ & $+0.18$ & $+0.31$ \\

$\langle$[Eu/Fe]$\rangle$ & $+0.25$ & $+0.15$ \\
\end{tabular}
%\end{minipage}
\end{table}

The sample in \cite{Wylie2006} show a general enhancement of the
{\it s}-process elemental abundances across the sample, with the
average enhancement for Zr being [Zr/Fe]~$=+0.65$~dex. A similar
abundance enhancement was found for several {\it s}-process elements
in \cite{Brown1992}, but Zr was found to be depleted with an average
abundance of [Zr/Fe]~$=-0.22$~dex. Y was enhanced in both studies as
was La and Eu.

The goal of this analysis of Lee~2525 is to link these two studies
in an effort to consolidate the reported abundances of heavy
elements in 47~Tuc giant branch stars.

\section{Spectral Derivation of Stellar Parameters}
\label{sec:SpectralDerivation} For sufficiently high resolution
observations the determination of the spectral parameters of a star
can be carried out using a curve of growth analysis. A curve of
growth analysis allows the simultaneous determination of the
effective temperature ($T_{eff}$), surface gravity ($log \ g$),
metallicity ([Fe/H]) and microturbulence ($\xi$) of a star. For both
Arcturus and Lee 2525, the normalised spectrum was measured for the
equivalent widths of selected Fe I and Fe II lines. Using MOOG
\citep{MOOG}, these equivalent widths were used to determine the
best fit stellar model by balancing the derived abundances
($log(\epsilon)$) with excitation potential ($\chi$) to find
$T_{eff}$, with reduced equivalent width ($log(\frac{W}{\lambda})$)
to find $\xi$, and with wavelength $\lambda$. The derived
$log(\epsilon)$ values for Fe I and Fe II were required to balance
in order to find the correct $log \ g$.

\subsection{Arcturus calibration}\label{sec:ArcCalib}

Deriving abundances relative to the Sun may not be appropriate for
giant stars as so many features present in giant stars are not
present in the solar spectrum. Deriving the abundances relative to
Arcturus, a giant star of similar metallicity to 47~Tuc, is a more
appropriate choice as the similarities in atmospheric structure will
cancel out any systematic errors \citep{Koch2008}.

The equivalent widths of spectral lines in Arcturus were measured in
order to calibrate the analysis process used in this study. The
equivalent widths of all possible Fe I and Fe II lines were measured
from the high resolution atlas of Arcturus \citep{Hinkle2005}.

Lines with equivalent width greater than 180m\AA~ were rejected as
they are saturated and also less accurate due to the assumption of a
gaussian profile fit in the determination of equivalent width. The
stellar parameters for Arcturus were derived using the MOOG curve of
growth analysis abfind function. The latest available published $log
\ gf$ values from VALD were used for this Fe line list
\citep{VALDKupka2000}.

MOOG derives abundances based on local thermodynamic equilibrium
calculations and assumes the model has a plane parallel geometry for
the structure of the stellar atmosphere. Spherical geometry stellar
models are more representative of the atmospheric structure of giant
stars and can be used in MOOG with the above codicil in mind. The
stellar parameters for Arcturus were derived using a MARCS spherical
geometry model \citep{Gustafsson2008} and an Kurucz/Atlas9 plane
parallel model (\citealt{Kurucz1979ModelGrid};
\citealt{Kurucz1995AtomMolBank}). Table~\ref{tab:FulbrightArcComp}
compares the Arcturus stellar parameters derived here using plane
parallel and spherical geometry with the values obtained in
\cite{Fulbright2006}.

\begin{table*}
\vspace{1cm}
\begin{minipage}{120mm}
\center \caption{Comparison of the derived light element abundances
for Arcturus from \citet{Fulbright07} and this study. In this study
two stellar models were considered in order to compare the results
from MOOG of spherical versus plane parallel geometry input models.}
\label{tab:FulbrightArcComp}
% Table generated by Excel2LaTeX from sheet 'FulArcAbunComp'
\begin{tabular}{cccccccccc}

 & \multicolumn{ 3}{c}{Fulbright06} &  & \multicolumn{ 5}{c}{This study} \\

 & \multicolumn{ 3}{c}{--------------------------} &  & \multicolumn{ 5}{c}{-----------------------------------------------} \\

Geometry & \multicolumn{ 2}{l}{Plane Parallel} &  &  & \multicolumn{ 2}{l}{Spherical} & \multicolumn{ 2}{l}{Plane Parallel} &  \\

$T_{eff}$ & \multicolumn{ 2}{l}{4290K} &  &  & \multicolumn{ 2}{l}{4300K} & \multicolumn{ 2}{l}{4270K} &  \\

$log \ g$ & \multicolumn{ 2}{l}{1.55} &  &  & \multicolumn{ 2}{l}{1.6} & \multicolumn{ 2}{l}{1.7} &  \\

[Fe/H] & \multicolumn{ 2}{l}{-0.5dex} &  &  & \multicolumn{ 2}{l}{-0.6dex} & \multicolumn{ 2}{l}{-0.60dex} &  \\

$\xi$ & \multicolumn{ 2}{l}{1.67$kms^{-1}$} &  &  & \multicolumn{ 2}{l}{1.50$kms^{-1}$} & \multicolumn{ 2}{l}{1.50kms$^{-1}$} &  \\

 &  &  &  &  &  &  &  &  &  \\

X & [X/Fe] & $\sigma$ & N &  & [X/Fe] & $\sigma$ & [X/Fe] & $\sigma$ & N \\

O & 0.48 & - & 1 &  & 0.57 & 0.02 & 0.56 & 0.03 & 2 \\

Na & 0.09 & - & 1 &  & 0.15 & 0.04 & 0.14 & 0.04 & 2 \\

Mg & 0.39 & 0.06 & 5 &  & 0.34 & 0.15 & 0.32 & 0.15 & 8 \\

Al & 0.38 & 0.03 & 3 &  & 0.25 & 0.07 & 0.24 & 0.07 & 4 \\

Si & 0.35 & 0.05 & 15 &  & 0.24 & 0.14 & 0.21 & 0.14 & 10 \\

Ca & 0.21 & 0.01 & 2 &  & 0.19 & 0.06 & 0.19 & 0.06 & 12 \\

Ti & 0.26 & 0.04 & 24 &  & 0.34 & 0.15 & 0.34 & 0.11 & 29 \\

 &  &  &  &  &  &  &  &  &  \\

 &  &  &  &  & [X/H] & $\sigma$ & [X/H] & $\sigma$ & N \\

Fe & - & - & - &  & -0.59 & 0.12 & -0.62 & 0.11 & 40 \\

\end{tabular}
\end{minipage}
\end{table*}

The best fitting stellar parameters for both the spherical and plane
parallel geometry models are in close agreement with each other and
with that derived in \cite{Fulbright2006}.

The determination of the remaining element abundances proceeded by
two methods. For the light elements (O through Zn), which have
sufficiently isolated lines, equivalent widths were measured and
used with the best fit model and {\it abfind} in MOOG to derive
abundances. For the weaker lines and those of the heavy elements (Y
to Eu) in more crowded spectral regions, abundances were derived by
comparing synthesised and observed spectra. The spectrum synthesis
line lists in key {\it s}-process regions were calibrated to the
observed Arcturus spectrum using {\it synth} in MOOG. The $log \ gf$
values for the key {\it s}-process lines were taken from the latest
published laboratory values: \cite{Lawler2001}, \cite{Biemont1981},
\cite{Den2003}, \cite{Hannaford1982}.

An abundance analysis of the light elements in Arcturus was carried
out in \cite{Fulbright07} and the values derived here are compared
with that study in Table~\ref{tab:FulbrightArcComp}. There is very
little change between the light element abundances derived using the
spherical model compared with the plane parallel model. Comparing
the spherical model to the \cite{Fulbright07} values there is
reasonable agreement ($\leq 0.1$dex) between the two sets of
abundances. Given the close agreement between the derived abundances
of the spherical and plane parallel models, and the good agreement
with \cite{Fulbright07}, the spherical model was selected as the
stellar model for Arcturus in this analysis.

The derived Arcturus abundances for the light and heavy species, and
uncertainties associated with changes in $T_{eff}$, $log \ g$ and
$\xi$, are listed in full in Table~\ref{tab:Arcturus_abundances}.

\begin{table}
%\vspace{1cm}
\begin{minipage}{80mm}
\centering \caption{Derived element abundances for Arcturus with
uncertainties in [Fe/H] and [X/Fe] associated with changes in
$T_{eff}$, $log \ g$ and $\xi$ using spherical geometry stellar
models.}\label{tab:Arcturus_abundances}
% Table generated by Excel2LaTeX from sheet 'ArcAbunErr'
\begin{tabular}{ccccccc}

 &  &  &  & \multicolumn{ 2}{r}{$\Delta$[X/H]} &  \\

\multicolumn{ 2}{l}{Species} &  &  & $\Delta T_{eff}$ & $\Delta log \ g$ & $\Delta \xi$ \\

X & N  & [X/H] & $\sigma$ & +100K & +0.5 & +0.5$kms^{-1}$ \\

 &  &  &  & \multicolumn{ 3}{c}{---------------------------------------} \\

Fe I & 29 & -0.61 & 0.12 & 0.06 & 0.13 & -0.29 \\

Fe II & 11 & -0.56 & 0.05 & -0.20 & 0.21 & -0.13 \\

 &  &  &  &  &  &  \\

X & N  & [X/Fe] & $\sigma$ & \multicolumn{ 2}{r}{$\Delta$[X/Fe]} &  \\

 &  &  &  & \multicolumn{ 3}{c}{---------------------------------------} \\

O I & 2 & 0.57 & 0.02 & -0.01 & 0.20 & -0.02 \\

Na I & 2 & 0.15 & 0.04 & 0.09 & 0.01 & -0.10 \\

Mg I & 8 & 0.34 & 0.15 & 0.01 & 0.03 & -0.04 \\

Al I & 4 & 0.25 & 0.07 & 0.07 & 0.01 & -0.07 \\

Si I & 10 & 0.24 & 0.14 & -0.06 & 0.13 & -0.06 \\

Ca I & 12 & 0.19 & 0.06 & 0.12 & -0.03 & -0.26 \\

Sc II & 2 & 0.24 & 0.01 & -0.02 & 0.22 & -0.18 \\

Ti I & 24 & 0.35 & 0.12 & 0.17 & 0.04 & -0.15 \\

Ti II & 5 & 0.33 & 0.10 & -0.04 & 0.22 & -0.15 \\

Zn I & 2 & -0.04 & 0.09 & -0.07 & 0.14 & -0.26 \\

 &  &  &  &  &  &  \\

Y I & 3 & 0.07 & 0.24 & 0.21 & 0.03 & -0.07 \\

Y II & 5 & 0.12 & 0.11 & 0.02 & 0.22 & -0.04 \\

Zr I & 7 & 0.01 & 0.07 & 0.13 & 0.07 & 0.00 \\

Zr II & 3 & 0.12 & 0.10 & -0.01 & 0.24 & -0.01 \\

Ba II & 2 & -0.19 & 0.08 & 0.27 & 0.26 & -0.16 \\

La II & 6 & 0.04 & 0.08 & 0.04 & 0.19 & -0.02 \\

Nd II & 4 & 0.10 & 0.07 & 0.03 & 0.17 & -0.09 \\

Eu II & 2 & 0.36 & 0.04 & -0.02 & 0.23 & -0.01 \\

\end{tabular}
\end{minipage}
\end{table}

The uncertainties associated with changes in $T_{eff}$, $log \ g$
and $\xi$ are sufficiently greater than the abundance differences
between the spherical and plane parallel models that the stellar
parameter selection can be considered to introduce a more
significant error than the geometry upon which the model is based.
The parameter uncertainties can account for differences in the light
element abundances between this study and \cite{Fulbright07}, except
for the Al I which is due to different line selection between the
studies. With regard to the heavy elements the neutral species are
most affected by changes in $T_{eff}$, while the ionised species are
more affected by changes in $log \ g$. The Ba II abundance varies
greatly with all three parameters which can be explained due to it
being derived from two strong lines that are sensitive to variations
with $\xi$ and $T_{eff}$, and, as an ionised species, a large
variation with $log \ g$.

\subsection{Lee 2525 stellar model}\label{sec:LeeStellarModel}
As part of an ongoing programme of research into the nature of {\it
s}-process elemental abundances in globular cluster giant stars the
Zr discrepancy between the studies of \cite{Brown1992} and
\cite{Wylie2006} needed to be addressed. This motivated the high
resolution observation and analysis of Lee 2525. The stellar
parameters derived for Lee 2525 in \cite{Brown1992} were
$T_{eff}=4225$~K, $log \ g=1.3$, [Fe/H]~$=-0.82$~dex and
$\xi=2.0$~kms$^{-1}$.

\begin{figure}
\begin{minipage}{85mm}
\begin{center}
\includegraphics[width = 80mm]{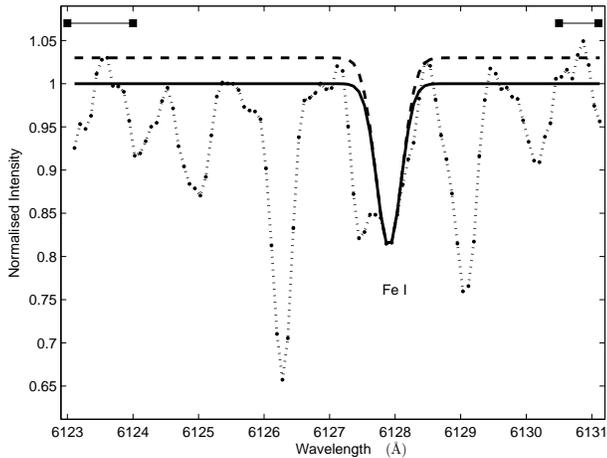}
\caption{Fe I line at 6127.906~\AA\ in the Lee 2525 stellar
spectrum. The dotted line is the observed spectrum. The black square
line segments are the continuum regions used to normalise the
spectrum locally about the line. The solid line is the line profile
used to determine the equivalent width for this line where the
continuum has been placed so as to account for the low SNR of the
spectrum. The dashed line is the line profile where the SNR has not
been taken into account.} \label{fig:FeEWContinuum}
\end{center}
\end{minipage}
\end{figure}

Equivalent widths from the Lee~2525 spectrum were measured for each
of the Fe I and Fe II lines that were used in the analysis of
Arcturus. Due to the much lower resolution and SNR of this spectrum
compared with the Arcturus atlas, careful selection of the best
fitted lines was made with which to determine the stellar parameters
for Lee~2525. This reduced the list from 40 to 22 lines in total.
Local normalisation about each Fe line in the stellar spectra was
necessary to ensure the correct location of the continuum.
\cite{Fulbright2006} defined continuum regions that were clear of
molecular bands about each of the Fe lines in Arcturus. These
regions were used in the local normalisation about each of the Fe
lines measured in Lee~2525. An example of the normalisation and
measurement of equivalent width process for the Fe I line at
6127.906~\AA\ in the Lee 2525 stellar spectrum is shown in
Figure~\ref{fig:FeEWContinuum}.

The normalisation was made by finding the mean intensity and
wavelength of each of the Fe lines' two continuum regions. The mean
was calculated iteratively, discarding points that lay outside
2$\sigma$ of the mean and then recalculating until less than 5\% of
the remaining points lay outside the 2$\sigma$ limit. A linear
relation between the two mean points was then divided out from the
spectrum resulting in the required normalisation. If only one
continuum region was available, the mean intensity of that region
was divided out of the spectrum to effect the normalisation. Given
the low SNR of the spectrum, this continuum placement was considered
to be too high (see dashed line profile in
Figure~\ref{fig:FeEWContinuum}). A further downward vertical shift
of 0.03 (as the noise per pixel was equivalent to $\pm3\%$ of the
signal) was applied in order to take account of the high degree of
noise in the spectrum (solid line profile in
Figure~\ref{fig:FeEWContinuum}).

Figure~\ref{fig:FeEWContinuum} clearly shows that a much larger
equivalent width would be measured if the noise was not taken into
account. For the Fe I line at 6127.906~\AA\ the equivalent width for
which the SNR was accounted for in the continuum placement was
measured to be 98.8~m\AA. If the SNR was not accounted for in the
continuum placement the equivalent width was measured to be
119.4~m\AA. Similar measurements were carried out for each of the Fe
lines. This shows that the placement of the continuum, and
consequently the measurement of the equivalent widths is a key
source of uncertainty in this analysis. The Fe and atomic lines were
carefully inspected and consistently measured by the above procedure
in order to minimise this uncertainty.

In this study the \cite{Brown1992} values of $T_{eff}$ and $log \ g$
(4225~K and 1.3 respectively) for Lee~2525 were used as the starting
point for determining the spectroscopic stellar model. An
exploration in stellar parameter space was carried out in order to
determine the best fit model. Table~\ref{tab:BWmodelComp} lists the
stellar parameters and resulting [Fe I,Fe II/H]\footnote{[Fe I,Fe
II/H] refers to the Fe abundance derived from Fe I and Fe II lines
respectively} values for each model permutation.

\begin{table*}
\begin{minipage}{160mm}
\center \caption{Stellar model parameters and resulting values of
[Fe I/H] and [Fe II/H] for Arcturus and range of models for
Lee~2525. Values at two different $\xi$ values are also compared.
The first model for Lee~2525 uses the \citet{Brown1992} parameters.
The final model for Lee~2525 is the best fit determined in this
study.} \label{tab:BWmodelComp}
% Table generated by Excel2LaTeX from sheet 'BWmodelcompAbundStd'
\begin{tabular}{crrrrrrrr}

Star & \multicolumn{ 2}{c}{Arcturus} & \multicolumn{ 6}{c}{Lee 2525} \\

 & \multicolumn{ 2}{l}{------------------} & \multicolumn{ 6}{c}{---------------------------------------------------} \\

$T_{eff}$ (K) & \multicolumn{ 2}{c}{4300} & \multicolumn{ 2}{c}{4225} & \multicolumn{ 2}{c}{4050} & \multicolumn{ 2}{c}{4225} \\

$log \ g$ & \multicolumn{ 2}{c}{1.6} & \multicolumn{ 2}{c}{1.3} & \multicolumn{ 2}{c}{0.8} & \multicolumn{ 2}{c}{1.2} \\

[Fe/H] (dex) & \multicolumn{ 2}{c}{$-0.60$} & \multicolumn{ 2}{c}{$-0.82$} & \multicolumn{ 2}{c}{$-0.65$} & \multicolumn{ 2}{c}{$-0.70$} \\

 &  &  &  &  &  &  &  &  \\

$\xi$ ($kms^{-1}$) & \multicolumn{ 2}{c}{1.5} & \multicolumn{ 2}{c}{1.5} & \multicolumn{ 2}{c}{1.5} & \multicolumn{ 2}{c}{1.8} \\

[Fe I/H] & $-0.61$ & $\pm$0.12 & $-0.53$ & $\pm$0.18 & $-0.64$ & $\pm$0.17 & $-0.72$ & $\pm$0.16 \\

[Fe II/H] & $-0.56$ & $\pm$0.05 & $-0.71$ & $\pm$0.05 & $-0.60$ & $\pm$0.02 & $-0.74$ & $\pm$0.08 \\

 &  &  &  &  &  &  &  &  \\

$\xi$ ($kms^{-1}$) & \multicolumn{ 2}{c}{2.0} & \multicolumn{ 2}{c}{2.0} & \multicolumn{ 2}{c}{2.0} & \multicolumn{ 2}{c}{2.0} \\

[Fe I/H] & $-0.90$ & $\pm$0.13 & $-0.86$ & $\pm$0.18 & $-0.98$ & $\pm$0.20 & $-0.84$ & $\pm$0.18 \\

[Fe II/H] & $-0.69$ & $\pm$0.10 & $-0.81$ & $\pm$0.09 & $-0.72$ & $\pm$0.06 & $-0.77$ & $\pm$0.09 \\

\end{tabular}
\end{minipage}
\end{table*}

\begin{figure*}
\begin{minipage}{150mm}
\begin{center}
\includegraphics[width = 150mm]{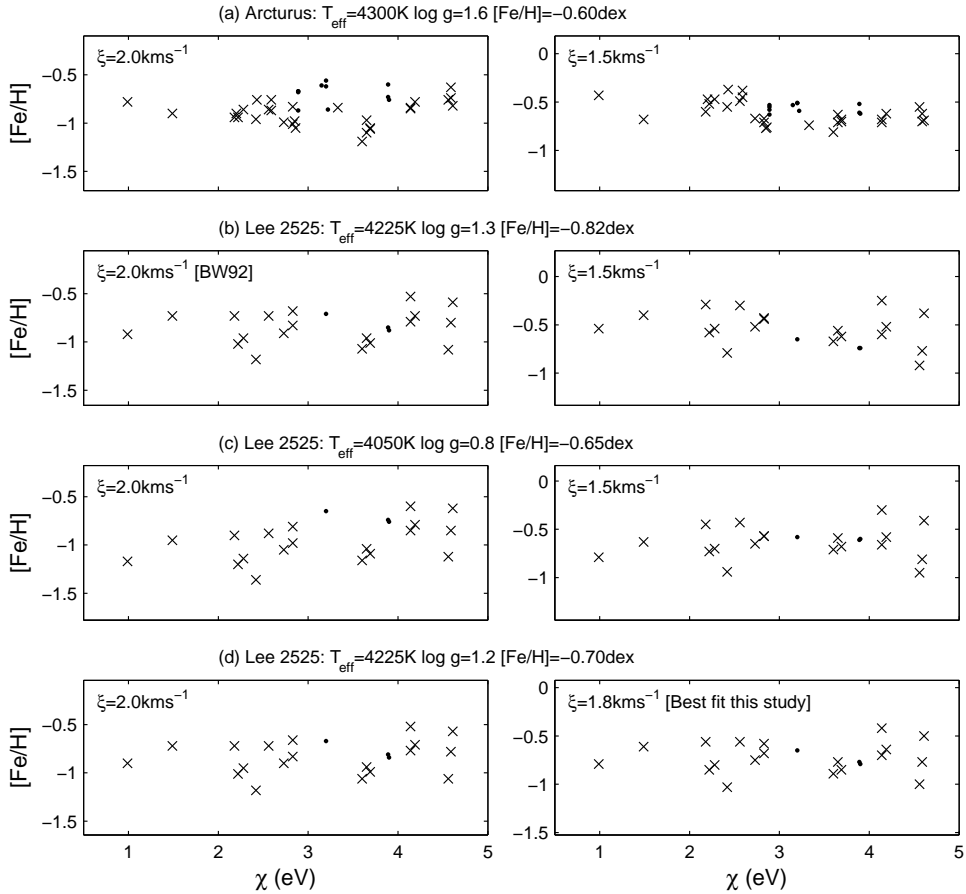}
\caption{Comparison of [Fe/H] vs $\chi$ derived from Fe I ($\times$)
and Fe II ($\bullet$) lines. (a) [Fe I/Fe II/H] derived for Arcturus
using the measured Arcturus equivalent widths. (b) As for (a) but
for the \citet{Brown1992} stellar model and the Lee~2525 measured
equivalent widths. (c) As for (b), but for a high metallicity model
and the Lee~2525 equivalent widths. (d) As for (b), but for the best
fit stellar model determined for Lee~2525 in this study and the
Lee~2525 equivalent widths. The figures in the lefthand column show
the derived [Fe I/Fe II/H] values for the specified stellar
atmosphere models at $\xi = 2.0$~kms$^{-1}$. The figures in the
righthand column are the same but for $\xi = 1.5$~kms$^{-1}$ except
for the best fit model in (d).} \label{fig:EPAbunBWArc}
\end{center}
\end{minipage}
\end{figure*}

Figure~\ref{fig:EPAbunBWArc} compares [Fe/H] against $\chi$ for
Arcturus, the \cite{Brown1992} model for Lee 2525 and two
prospective models for Lee 2525 derived in this study.

In Figure~\ref{fig:EPAbunBWArc}a the small spread in values is
obvious for Arcturus due to the much greater resolution and high
signal to noise of that spectrum. The Arcturus models at
$\xi=2.0$~kms$^{-1}$ and $\xi=1.5$~kms$^{-1}$ are compared showing
that Arcturus clearly falls in the $\xi=1.5$~kms$^{-1}$ regime,
illustrated by the smaller spread in [Fe/H] values (see
Table~\ref{tab:BWmodelComp}).

The stellar model matching the \cite{Brown1992} parameters for
Lee~2525 (Figure~\ref{fig:EPAbunBWArc}b) show good agreement between
the derived Fe I and Fe II abundances in the $\xi=2.0$~kms$^{-1}$
regime. For the $\xi=1.5$~kms$^{-1}$ regime the [Fe I,Fe II/H]
values are out of equilibrium and the [Fe I/H] disagrees with the
model [Fe/H] by $\sim0.2$~dex (see Table~\ref{tab:BWmodelComp}).
Figure~\ref{fig:EPAbunBWArc}c uses a model for Lee 2525 at
[Fe/H]~$=-0.65$~dex which returns [Fe/H]~$=-0.62$~dex for
$\xi=1.5$~kms$^{-1}$. The spread in values are reasonable and
tighter than the spread in the $\xi=2.0$~kms$^{-1}$ regime.

Ultimately the best fit model, shown in
Figure~\ref{fig:EPAbunBWArc}d resides closer to the
$\xi=2.0$~kms$^{-1}$ regime at $\xi=1.8$~kms$^{-1}$ where the spread
in values is reasonable for both the Fe I and Fe II abundances.
Hence the best fit model for Lee~2525 derived in this study returned
values of $T_{eff}=4225$~K, $log \ g=1.2$, [Fe/H]~$=-0.70$~dex and
$\xi =1.8$~kms$^{-1}$ (see Table~\ref{tab:BWmodelComp}), which are
in reasonable agreement with the values derived in \cite{Brown1992}.

The Fe abundances derived from the Fe I and Fe II lines using the
best fit stellar atmosphere model were $-0.72\pm0.16$~dex and
$-0.74\pm0.08$~dex respectively. These values were derived from the
equivalent widths measured such that the continuum placement took
account of the low SNR of the Lee 2525 spectrum, as outlined above.
Using the best fit Lee 2525 stellar model, Fe abundances were
re-derived from the equivalent widths measured without taking the
SNR into account. The respective values were determined to be [Fe
I/H]~$-0.37\pm0.17$~dex and $-0.29\pm0.10$~dex. These values are
considerably more metal-rich and reflect an increase in derived
abundance of $\sim 0.3$~dex from the Fe I lines and $\sim 0.4$~dex
from the Fe II lines due to the increased equivalent width values.
The weaker Fe II lines show a larger change as the increase in
equivalent width is proportionally greater than for the strong Fe I
lines. These values reflect the maximum possible uncertainty
introduced by the spectrum's low SNR. However, the equivalent width
measurements and abundance analysis were carried out in a consistent
manner with careful inspection of the lines to ensure the best
placement of the continuum in order to minimise this uncertainty.

The best fit stellar model for this study returned an average [Fe/H]
of $-0.73$~dex which is slightly more metal-rich than the
\cite{Brown1992} value of $-0.82$~dex. Comparing these values to
previous studies of RGB and AGB stars in 47~Tuc, both are more
metal-poor than \cite{Alves-Brito2005}, which found an average
[Fe/H] of $-0.68$~dex, \cite{Carretta2004} ([Fe/H]~$=-0.67$~dex) and
\cite{Wylie2006} ([Fe/H]~$=-0.60$~dex). However, a more recent
paper, \cite{Koch2008}, derived an [Fe/H] of $-0.76$~dex for 47 Tuc
with their stars having a range of values from $-0.82$~dex to
$-0.72$~dex. This is significantly more metal-poor than previous
studies and in better agreement with the derived metallicities of
Lee 2525 found in this study and \cite{Brown1992}.

\section{Element abundances in Lee 2525}
Table~\ref{tab:L2525_derived_abundances} lists the light and heavy
elemental abundances derived in this study for two models for Lee
2525 as well as the results from \cite{Brown1992}. As outlined in
Section~\ref{sec:ArcCalib}, the light elemental abundances were
derived using {\it abfind} in MOOG, while heavy elemental abundances
were derived using MOOG's spectrum synthesis function, {\it synth}.

\begin{table*}
\begin{minipage}{160mm}
\center \caption{The light and heavy elemental abundances derived in
this study from two stellar models for Lee 2525 compared with the
values derived in \citet{Brown1992}. The uncertainty on the mean
($\sigma$) and number of lines used to derived the abundance (N) are
included. The variation in abundances due to changes in stellar
parameters corresponding to $\Delta T_{eff}=+100$~K, $\Delta log \
g=+0.5$ and $\Delta \xi=+0.5$~kms$^{-1}$ are also shown.}
\label{tab:L2525_derived_abundances}
%\begin{flushleft}
%\begin{minipage}{120mm}
% Table generated by Excel2LaTeX from sheet 'BWAbun'
\begin{tabular}{cccccccccccccc}

 & \multicolumn{ 3}{c}{BW92} &  & \multicolumn{ 5}{c}{This Study} &  &  &  &  \\

 & \multicolumn{ 3}{c}{---------------------------} &  & \multicolumn{ 5}{c}{-----------------------------------------------} &  &  &  &  \\

$T_{eff}$ (K) & 4225 &  &  &  & 4225 &  & 4225 &  &  &  &  &  &  \\

$log \ g$ & 1.3 &  &  &  & 1.3 &  & 1.2 &  &  &  &  &  &  \\

[Fe/H] (dex) & -0.82 &  &  &  & -0.82 &  & -0.70 &  &  &  &  &  &  \\

$\xi (kms^{-1})$ & 2.0 &  &  &  & 2.0 &  & 1.8 &  &  &  & \multicolumn{ 2}{r}{$\Delta$[X/H]} &  \\

 &  &  &  &  &  &  &  &  &  &  & $\Delta T_{eff}$ & $\Delta log \ g$ & $\Delta \xi$ \\

Species & [X/H] & $\sigma$ & N &  & [X/H] & $\sigma$ & [X/H] & $\sigma$ & N &  & +100K & +0.5 & +0.5$kms^{-1}$ \\

 &  &  &  &  &  &  &  &  &  &  & \multicolumn{ 3}{l}{--------------------------------------------} \\

Fe I & -0.83 & 0.16 & 22 &  & -0.85 & 0.18 & -0.72 & 0.16 & 19 &  & 0.01 & 0.11 & -0.23 \\

Fe II & -0.82 & 0.17 & 5 &  & -0.81 & 0.09 & -0.73 & 0.08 & 3 &  & -0.12 & 0.32 & -0.06 \\

 &  &  &  &  &  &  &  &  &  &  &  &  &  \\

 & [X/Fe] & $\sigma$ & N &  & [X/Fe] & $\sigma$ & [X/Fe] & $\sigma$ & N &  & \multicolumn{ 2}{r}{$\Delta$[X/Fe]} &  \\

 &  &  &  &  &  &  &  &  &  &  & \multicolumn{ 3}{c}{--------------------------------------------} \\

O I & - & - & 0 &  & 0.45 & - & 0.40 & - & 1 &  & 0.01 & 0.18 & -0.01 \\

Na I & -0.01 & 0.06 & 2 &  & 0.15 & 0.06 & 0.05 & 0.05 & 2 &  & 0.09 & 0.01 & -0.07 \\

Mg I & - & - & 0 &  & 0.43 & 0.02 & 0.34 & 0.04 & 2 &  & 0.01 & 0.05 & -0.06 \\

Al I & 0.51 & 0.23 & 2 &  & 0.32 & 0.02 & 0.21 & 0.03 & 2 &  & 0.08 & 0.02 & -0.08 \\

Si I & 0.21 & 0.14 & 6 &  & 0.43 & 0.20 & 0.36 & 0.20 & 7 &  & -0.07 & 0.14 & -0.06 \\

Ca I & -0.03 & 0.06 & 9 &  & 0.04 & 0.21 & 0.00 & 0.23 & 8 &  & 0.11 & 0.00 & -0.24 \\

Sc II & -0.04 & 0.06 & 2 &  & 0.06 & 0.07 & 0.05 & 0.08 & 2 &  & -0.03 & 0.22 & -0.13 \\

Ti I & 0.17 & 0.09 & 7 &  & 0.24 & 0.19 & 0.14 & 0.18 & 9 &  & 0.17 & 0.05 & -0.09 \\

Ti II & 0.42 & - & 1 &  & 0.19 & 0.06 & 0.17 & 0.04 & 2 &  & -0.05 & 0.23 & -0.11 \\

Zn I & 0.14 & 0.03 & 2 &  & 0.04 & 0.15 & 0.05 & 0.16 & 2 &  & -0.08 & 0.16 & -0.24 \\

 &  &  &  &  &  &  &  &  &  &  &  &  &  \\

Y I & 0.58 & - & 1 &  & 0.60 & 0.25 & 0.51 & 0.28 & 2 &  & 0.25 & 0.09 & 0.01 \\

Y II & 0.41 & 0.47 & 2 &  & 0.44 & - & 0.54 & - & 1 &  & 0.07 & 0.20 & -0.50 \\

Zr I & -0.51 & 0.24 & 3 &  & 0.37 & 0.04 & 0.36 & 0.04 & 3 &  & 0.15 & 0.03 & -0.12 \\

Zr II & - & - & 0 &  & 0.70 & - & 0.65 & - & 1 &  & 0.05 & 0.25 & -0.03 \\

Ba II & -0.15 & 0.21 & 3 &  & -0.32 & 0.02 & -0.21 & 0.01 & 2 &  & 0.06 & 0.27 & -0.49 \\

La II & 0.10 & 0.40 & 2 &  & 0.34 & 0.01 & 0.31 & 0.03 & 3 &  & 0.04 & 0.24 & -0.05 \\

Nd II & - & - & 0 &  & 0.40 & - & 0.41 & - & 1 &  & 0.11 & 0.31 & 0.05 \\

Eu II & 0.44 & - & 1 &  & 0.48 & - & 0.40 & - & 1 &  & 0.12 & 0.38 & 0.11 \\

 &  &  &  &  &  &  &  &  &  &  &  &  &  \\

 & [X/Y] & $\sigma$ &  &  & [X/Y] & $\sigma$ & [X/Y] & $\sigma$ &  &  & \multicolumn{ 2}{r}{$\Delta$[X/Y]} &  \\

 &  &  &  &  &  &  &  &  &  &  & \multicolumn{ 3}{c}{--------------------------------------------} \\

{\it ls}/Fe & 0.16 & 0.50 &  &  & 0.53 & 0.20 & 0.51 & 0.22 &  &  & 0.13 & 0.14 & -0.16 \\

{\it hs}/Fe & 0.10 & 0.40 &  &  & 0.37 & 0.01 & 0.36 & 0.03 &  &  & 0.08 & 0.27 & 0.00 \\

{\it hs/ls} & -0.06 & 0.64 &  &  & -0.16 & 0.15 & -0.16 & 0.18 &  &  & -0.05 & 0.13 & 0.16 \\

\end{tabular}
\end{minipage}
%\end{flushleft}
\end{table*}

Given the similar nature of the stellar parameters for Lee~2525
derived in this study to those derived in \cite{Brown1992},
elemental abundances using both models were calculated for a
complete comparison. In Table~\ref{tab:L2525_derived_abundances} the
abundances derived in \cite{Brown1992} are quoted in column 2 with
associated uncertainties and the number of lines used. Column 5
lists the abundances derived in this study using the
\cite{Brown1992} stellar model parameters, and column 7 lists the
abundances derived using the best fit model determined in this
study. Columns 10 to 12 list the changes in abundance with
associated changes in $T_{eff}$, $log \ g$ and $\xi$.

As described in Section~\ref{sec:LeeStellarModel} the placement of
the continuum for the measurement of the equivalent widths
introduced a key source of uncertainty for the abundances measured
in this analysis. This is likely to be the main source of
discrepancy between this analysis and the element abundances
determined in \cite{Brown1992}. While the spectra in both studies
was of similar resolution ($R\sim20,000$) the SNR in this study was
considerably lower. It must also be noted that \cite{Brown1992} used
a mixture of solar and laboratory {\it log gf} values but only
laboratory {\it log gf} values were used in this analysis. Also
\cite{Brown1992} used the solar abundances reported in
\cite{Anders1989} while those of \cite{Lodders2003} were used here.

\subsection{Light elements: O to Zn}\label{sec:light_elements}
Comparing the abundances we derived using the \cite{Brown1992}
parameters with the \cite{Brown1992} results, there is reasonable
agreement (to within $1 \sigma$) for the majority of the light
elements. The key differences were: Al~I, which was less abundant in
this study by 0.19~dex; Ti~II, which was less abundant by 0.23~dex;
Zn~I, which was less abundant by 0.10~dex; and Sc~II, which was over
abundant by 0.10~dex.

Comparing the best fit model of this study with \cite{Brown1992},
similar comments can be made. In general the light elemental
abundances agree within $1 \sigma$. Sc~II and Zn~I are also enhanced
using this model compared to \cite{Brown1992}, while Ti~II is still
significantly depleted. However, of the three sets of stellar
atmosphere models, the best fit model derived in this study provided
the best agreement between Ti~I and Ti~II abundances indicating a
better choice of {\it log g}, at least in terms of Ti.

The error analysis in Table~\ref{tab:L2525_derived_abundances} shows
that the strong lines of Sc~II and Zn~I used in this study are
highly sensitive to changes in $\xi$ as is the case of strong lines.
Also both Sc~II and Ti~II are sensitive to changes in gravity as is
expected for ionised lines.

This study confirms the abundance correlations previously observed
for this star in \cite{Brown1992}, namely a correlation of Al and Na
abundance with CN strength. Another key abundance anomaly is the
Na-O anti-correlation observed in 47~Tuc (see \cite{Carretta2004}
and references therein). An O abundance was not measured in
\cite{Brown1992}. However, it was obtained in this study using the
forbidden O I line at 6300~\AA. Lee 2525 is enhanced in O
([O/Fe]~$=0.40$~dex), while Na is not ([Na/Fe]~$=0.05\pm0.05$~dex).
These values fall clearly within the anti-correlated trend of
[Na/Fe] to [O/Fe] shown in Figure 5 of \cite{Carretta2004}.

The abundances of Mg and Al are both enhanced in agreement with
previous studies. There is no indication of an anti-correlation
between Mg and Al (\citealt{Carretta2004}; \citealt{Koch2008}). As
47~Tuc is a metal-rich globular cluster this anomaly is not expected
\citep{Gratton2004}.

With regard to Ca, previous studies have shown enhancements in
47~Tuc stars (\citealt{Carretta2004}; \citealt{Koch2008}. However,
the analysis of this star found a Ca abundance of
[Ca/Fe]~$=0.00$~dex, in agreement with the value determined in
\cite{Brown1992} ([Ca/Fe]~$=-0.03$~dex).

The Ti abundance is slightly enhanced ($\sim +0.15$~dex) in this
study for both Ti I and Ti II. In \cite{Brown1992} Ti I was enhanced
at this same level while Ti II was greatly enhanced ($\sim
+0.42$~dex). This implies that there is a {\it log g} determination
issue in their study due to the neutral and ionised species being
out of equilibrium.

For a sample of five red giants, \cite{Alves-Brito2005} found the
mean abundances for Ca to be [Ca/Fe]~$=0.0$~dex and Ti to be
[Ti/Fe]~$\sim 0.25$~dex. The Ca and Ti abundances derived in this
study agree with \cite{Alves-Brito2005} within the uncertainties.

In \cite{Alves-Brito2005} a comparison of Ca abundances was made
between \cite{Brown1992} and \cite{Carretta2004}. The lack of
enhancement in Ca in \cite{Brown1992} was noted. The stellar sample
in \cite{Carretta2004} all had enhancements in Na and Ca, while the
\cite{Alves-Brito2005} sample showed no enhancements in the mean
abundances of both Na and Ca. While the correlation of Na with CN
can be explained as leakage from the CNO cycle, a variation of Ca
with Na or CN is not expected. Further investigation of Ca
abundances is needed in order to clarify the reported abundances for
this element in 47 Tuc.

Overall this analysis of Lee confirms previous abundance anomalies
within 47 Tuc giant branch stars with regards to CN, Na and O.

\subsection{Heavy elements: Y to Eu}\label{sec:heavy_elements}
The comparison of the heavy elemental abundances show that the
current analysis of Lee~2525 carried out using the \cite{Brown1992}
stellar atmosphere model returns slightly higher abundances than the
values found in the paper itself. Zr and La show enhancements
relative to \cite{Brown1992}, while Ba is depleted.

The best fit stellar model in this study also shows Zr and La are
enhanced compared with the \cite{Brown1992} values. However, there
is better agreement between the Y I values and the Y II abundance
values. There is also a large uncertainty in the \cite{Brown1992} La
abundance within which the value from this study resides. While Ba
is depleted here with respect to \cite{Brown1992} the effect is not
so great and the abundances agrees within the uncertainties. While
the Eu abundance for both studies is based on only one line so no
$\sigma$ is available, the difference between the abundances is only
0.04dex. Given the high sensitivity of Ba II to $log \ g$  and $\xi$
it was not included here in the definition of the heavy {\it
s}-process index, {\it hs}. As such for this study {\it ls}~$=
\langle$Y~I,Y~II,Zr~I,Zr~II$\rangle$ and {\it
hs}~$=\langle$La~II,Nd~II$\rangle$.

\begin{figure*}
\begin{minipage}{160mm}
\begin{center}
\includegraphics[width = 160mm]{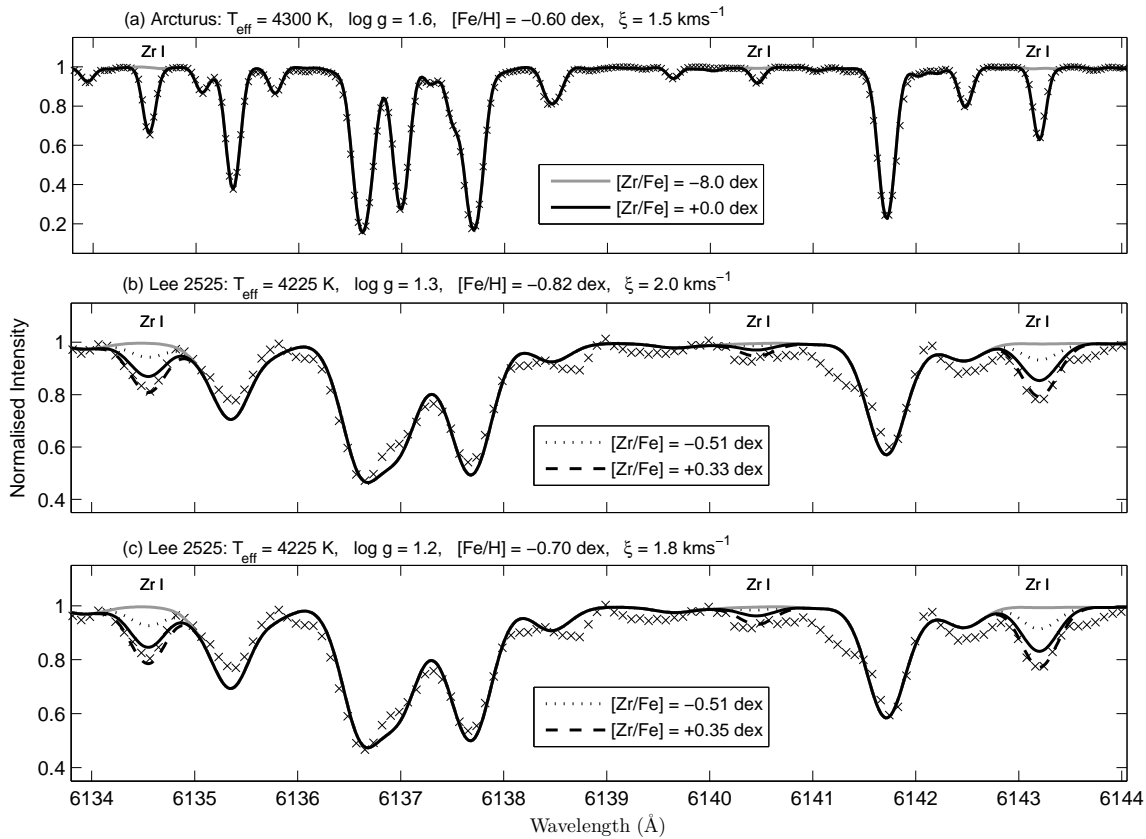}
\caption{Wavelength region about 6140~\AA\ showing the observed
spectrum overlaid with spectra synthesised at different specified Zr
abundances: (a) Arcturus spectrum where [Zr/Fe]~$=~+0.00$~dex
provides the best fit to the Zr I lines; (b) Lee 2525 spectrum using
the stellar model based on the stellar parameters specified in
\citet{Brown1992}. [Zr/Fe]~$=-0.51$~dex is the best fit Zr abundance
found in \citet{Brown1992}. [Zr/Fe]~$=+0.33$~dex is the best fit Zr
abundance for this model in the current analysis; (c) Lee 2525
spectrum using the best fit stellar model parameters determined in
this study. [Zr/Fe]~$=+0.35$~dex is the best fit to the Zr
abundance.} \label{fig:Zr6140comparison}
\end{center}
\end{minipage}
\end{figure*}

The Zr abundances are of particular interest as there is a
significant difference between them and the \cite{Brown1992}
results. Using the \cite{Brown1992} parameters an enhancement in Zr
of [Zr~I/Fe]~$=+0.37$~dex and [Zr~II/Fe]~$=+0.70$~dex was determined
in this study. The best fitting model returned values of
[Zr~I/Fe]~$=+0.36$~dex and [Zr~II/Fe]~$=+0.65$~dex compared with
[Zr~I/Fe]~$=-0.51$~dex as determined in \cite{Brown1992}. In order
to investigate these differences further
Figure~\ref{fig:Zr6140comparison} shows the observed spectrum for
Arcturus and Lee 2525 in a key region containing three Zr I lines
used in this study, two of which were used in \cite{Brown1992}.

In Figure~\ref{fig:Zr6140comparison}b Lee~2525 is overlaid with a
synthetic spectrum generated using the \cite{Brown1992} model. The
Zr abundance has been varied to show the spectrum when Zr is
depleted ([Zr/Fe]~$=-8.0$~dex), Zr abundance at the value calculated
in \cite{Brown1992} ([Zr/Fe]~$=-0.51$~dex), the Zr abundance
expected from the model ([Zr/Fe]~$=0.0$~dex) and the best fit Zr
abundance in this study for the line at 6134.57~\AA\
([Zr/Fe]~$=+0.33$~dex). Similarly in
Figure~\ref{fig:Zr6140comparison}c, a synthesised spectrum of the
best fit model in this study is overlaid on Lee~2525 with the same
variations to Zr, except the best fit Zr abundance for 6134.57~\AA\
is [Zr/Fe]~$=+0.35$~dex. In both cases the Zr features are enhanced
with respect to the model abundance confirming the expected result
of enhancements in Y coinciding with enhancements in Zr.

With Zr now similar to Y, these heavy elemental abundances fall
within the spread of values determined for the {\it s}-process
elemental abundances obtained in \cite{Wylie2006} confirming an
enhancement of {\it s}-process elements in 47~Tuc (see
Table~\ref{tab:Wylie_Browncomp}).

\subsection{Lee 2525 abundances with respect to Arcturus}\label{sec:Lee2525wrtArc}
In light of this result for Lee 2525 the [Fe/H] and [X/Fe] relative
to the Sun listed in Table~\ref{tab:L2525_derived_abundances} were
recalculated relative to Arcturus and are listed in
Table~\ref{tab:LeeAbundancesArc}. This was carried out in order to
remove any systematic errors within the analysis process by
re-stating the abundances in Lee 2525 relative to a star of similar
metallicity and atmospheric structure on which the process has been
calibrated \citep{Koch2008}.

\begin{table}
\caption{ [Fe/H], [X/Fe] and {\it s}-process abundance ratios for
Arcturus, and [Fe/H], [X/Fe] and {\it s}-process abundance ratios
calculated relative to Arcturus for the \citet{Brown1992} and best
fit Lee~2525  stellar models, where
[X/Y]$_{Arc}$=[X/Y]$_{\star}$-[X/Y]$_{Arcturus}$. The results from
\citet{Brown1992} are also re-calculated relative to Arcturus for
comparison.} \label{tab:LeeAbundancesArc}
% Table generated by Excel2LaTeX from sheet 'BWAbunwrtArc'
\begin{tabular}{ccccccc}

 &  &  &  &  & \multicolumn{ 2}{c}{This study} \\

 &  &  &  &  & \multicolumn{ 2}{c}{---------------------} \\

 & Arcturus &  & BW92 &  & BW & Best Fit \\

X & [X/H] &  & [X/H]$_{Arc}$ &  & [X/H]$_{Arc}$ & [X/H]$_{Arc}$ \\

Fe I & -0.61 &  & -0.22 &  & -0.24 & -0.11 \\

Fe II & -0.56 &  & -0.26 &  & -0.22 & -0.17 \\

 &  &  &  &  &  &  \\

 & [X/Fe] &  & [X/Fe]$_{Arc}$ &  & [X/Fe]$_{Arc}$ & [X/Fe]$_{Arc}$ \\

O I & 0.57 &  & - &  & -0.12 & -0.17 \\

Na I & 0.15 &  & -0.16 &  & 0.01 & -0.10 \\

Mg I & 0.34 &  & - &  & 0.08 & -0.01 \\

Al I & 0.25 &  & 0.26 &  & 0.06 & -0.04 \\

Si I & 0.20 &  & 0.01 &  & 0.23 & 0.16 \\

Ca I & 0.19 &  & -0.22 &  & -0.15 & -0.19 \\

Sc I & 0.24 &  & -0.28 &  & -0.18 & -0.20 \\

Ti I & 0.35 &  & -0.18 &  & -0.11 & -0.21 \\

Ti II & 0.33 &  & 0.09 &  & -0.14 & -0.16 \\

Zn I & -0.04 &  & 0.18 &  & 0.07 & 0.09 \\

 &  &  &  &  &  &  \\

Y I & 0.07 &  & 0.51 &  & 0.53 & 0.44 \\

Y II & 0.12 &  & 0.29 &  & 0.32 & 0.42 \\

Zr I & 0.01 &  & -0.52 &  & 0.36 & 0.35 \\

Zr II & 0.12 &  & - &  & 0.58 & 0.53 \\

Ba II & -0.19 &  & 0.04 &  & -0.13 & -0.02 \\

La II & 0.04 &  & 0.06 &  & 0.30 & 0.27 \\

Nd II & 0.10 &  & - &  & 0.31 & 0.32 \\

Eu II & 0.36 &  & 0.08 &  & 0.12 & 0.04 \\

 &  &  &  &  &  &  \\

 & [X/Y] &  & [X/Y]$_{Arc}$ &  & [X/Y]$_{Arc}$ & [X/Y]$_{Arc}$ \\

{\it ls}/Fe & 0.08 &  & 0.09 &  & 0.45 & 0.44 \\

{\it hs}/Fe & 0.07 &  & 0.06 &  & 0.30 & 0.29 \\

{\it hs/ls} & -0.01 &  & -0.03 &  & -0.14 & -0.14 \\

\end{tabular}
\end{table}

While Arcturus and Lee 2525 have similar stellar parameters,
Table~\ref{tab:LeeAbundancesArc} shows the clear differences in
their chemical make up. For the most part, Lee~2525 is less abundant
in the light elements than Arcturus and more abundant in the heavy
elements. Si and Zn are both enhanced in Lee 2525 compared to
Arcturus. However the derived uncertainties of Si and Zn in this
study bring them in line with the other light element abundances.

Of the heavy elements, Ba II and Eu II show abundances similar to
Arcturus. The strength of the Ba II lines and their sensitivity to
$\xi$ make the Ba II less reliable. As Eu is predominantly an {\it
r}-process element, its abundance is a useful indication of how much
pollution by supernov\ae\ the gas clouds underwent prior to the
formation of stars. The similar value between Lee 2525 and Arcturus
could imply some similar degree of exposure. The similarity in Eu
abundances between globular cluster stars and field stars has been
noted previously \citep{James2004b}.

The comparison between the Lee 2525 and Arcturus abundance results
is a natural consequence of undertaking a differential analysis with
a standard star. However, as Lee~2525 is a globular cluster star and
Arcturus is field star, the comparison being made is really between
two distinct stellar environments which is beyond the scope of this
paper. The key adjustments due to differencing Lee~2525 with
Arcturus is a reduction in the scatter of the {\it s}-process
element abundances. For Zr and Y there is better agreement between
their neutral and ionised abundances, and better agreement between
these two light {\it s}-process peak elements overall. The
abundances of the heavy {\it s}-process elements (La and Nd) are
also brought in line. These effects are reflected in the [{\it
hs}/Fe] and [{\it ls}/Fe] ratios although the [{\it hs/ls}] ratio is
only adjusted by $0.02$~dex. These improvements imply the reduction
of systematic errors in the results due to the analysis process. In
the case of a larger set of stars within a globular cluster a
similar differential analysis would be useful to reduce systematic
errors within a study and thereby to produce more consistent
results.

\subsection{The {\it hs} to {\it ls} ratio in 47 Tuc giant stars}
\begin{figure}
\begin{minipage}{80mm}
\includegraphics[width = 80mm]{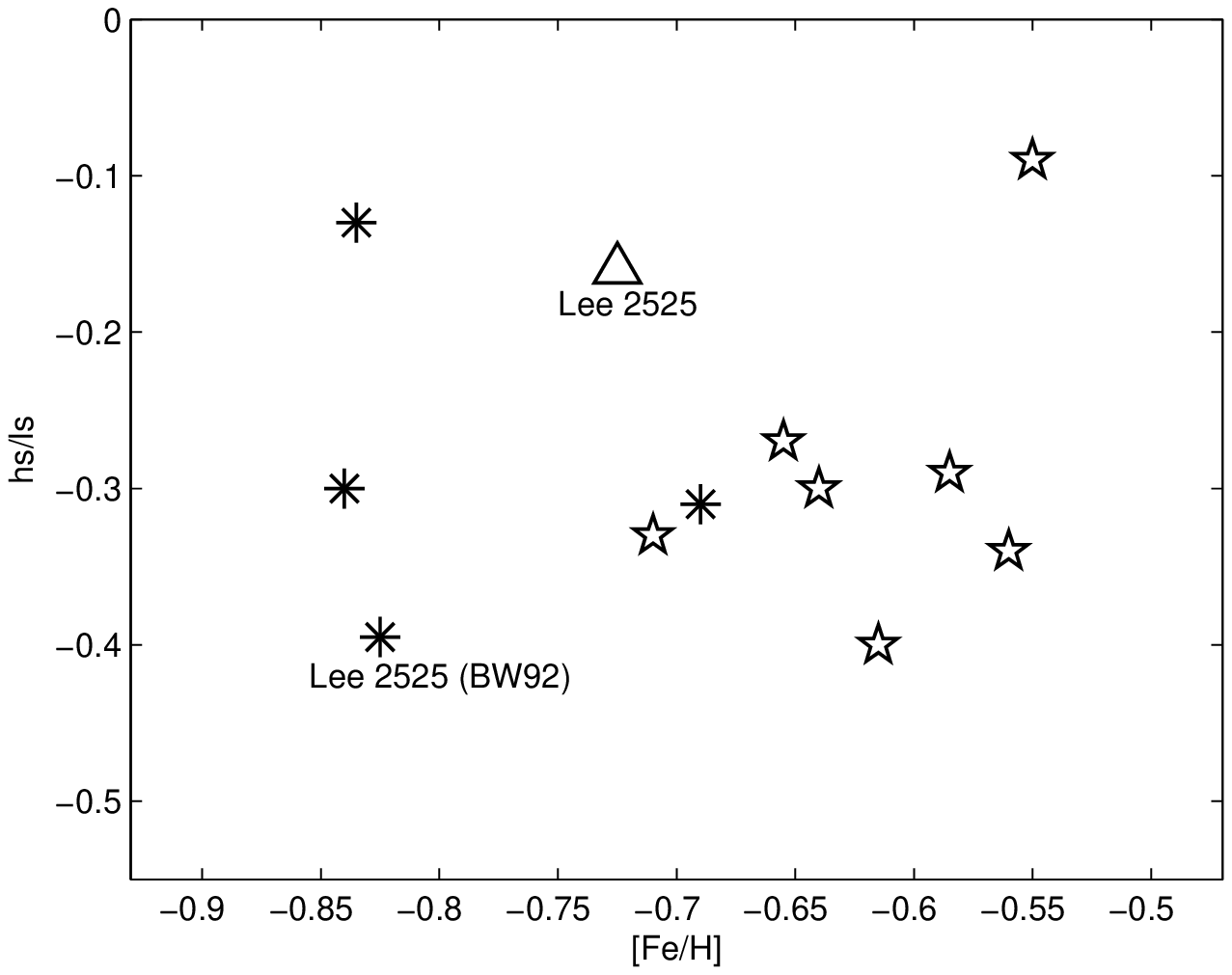}
\caption{The ratio of the heavy to light {\it s}-process element
abundance ([{\it hs/ls}]) for each star against [Fe/H]. ($\star$)
\citet{Wylie2006}; ($\ast$) \citet{Brown1992} using {\it ls} =
$\langle$Y I, Y II$\rangle$; and ($\bigtriangleup$) Lee 2525, this
study.} \label{fig:hslsVsFeH}
%\end{center}
\end{minipage}
\end{figure}

The {\it hs} and {\it ls} indices derived for Lee 2525 in this study
were compared with \cite{Brown1992} and \cite{Wylie2006}.
Figure~\ref{fig:hslsVsFeH} shows the abundance ratio of the heavy
{\it s}-process to light {\it s}-process elements ([{\it hs/ls}])
for Lee 2525 as well as for the other studies.

The {\it ls} indices for the \cite{Brown1992} stars were
re-calculated to use only the Y I and Y II abundances in light of
the above analysis of Zr. Considering the sample as a whole there
appears to be no trend between the [{\it hs/ls}] ratio and [Fe/H] in
Figure~\ref{fig:hslsVsFeH}. Given these stars all exist in the same
cluster and therefore have the same metallicity within some
variance, if there was a spread in [{\it hs/ls}] values greater than
the systematic spread on the [Fe/H] values, it would indicate that
{\it s}-process elements were being produced in these stars.

Observational and theoretical studies have shown that for AGB stars
undergoing third dredge up the [{\it hs/ls}] ratio generally
increases with decreasing metallicity, although the theoretical
relations are naturally more complex \citep{Busso2001}. As the seed
nuclei become fewer there is greater enhancement in the heavy {\it
s}-process peak compared with the light {\it s}-process peak. As no
such trend exists in this sample of 47 Tuc stars, it implies that
the {\it s}-process abundances observed here are primordial and are
not being produced internally in these stars.

\section{Conclusion}
Studies to date of {\it s}-process element abundances in 47 Tuc
stars have concluded that the observed {\it s}-process element
abundances are due to the primordial chemical composition of the
cluster or some pollution event early in the cluster's history.
While within each study there is agreement as to the magnitude of
the {\it s}-process element abundances, the abundances differ
between the studies. The analysis of Lee 2525 carried out here
resolves a discrepancy between \cite{Brown1992} and \cite{Wylie2006}
as to the magnitude of the Zr abundance in this star. The current
study found Lee 2525 to be enhanced in Zr at [Zr/Fe]~$=+0.51$~dex
which is in agreement with the enhancement found for Y of
[Y/Fe]~$=+0.53$~dex, another light {\it s}-process element. This is
in line with {\it s}-process element enhancements found in 47 Tuc
giant branch stars in \cite{Wylie2006}. The Na-CN correlation
reported in \cite{Brown1992} for Lee 2525 was also found here, as
well as an Na-O anti-correlation in line with other studies of light
elements in 47 Tuc stars \citep{Carretta2004}. These results support
the premise that the abundance anomalies observed in 47 Tuc have a
primordial or pollution-based origin.

While this study has resolved a discrepancy between two key papers
the overall {\it s}-process element abundance distribution within 47
Tuc is still not clear. It is necessary to expand the sample of 47
Tuc stars analysed for their {\it s}-process element abundances in
order to consolidate the results found here with those from other
studies. This analysis was carried out differentially with respect
to Arcturus in an effort to reduce systematic errors by calibrating
the analysis process to a standard star of similar stellar
parameters. This provides a solid framework for further study within
this area.

\section*{Acknowledgements}
The authors wish to acknowledge financial support from the New
Zealand Marsden Fund and the University of Canterbury, Christchurch,
New Zealand.

This research has made use of the SIMBAD database, operated at CDS,
Strasbourg, France. It has also made use of NASA's Astrophysics Data
System.

%\bsp

\label{lastpage}
%Included for Gather Purpose only:
%input "D:\Swapspace\Publications\MNRAS_Worley\Brown\ReSubmitted\worleyreferences.bib"
%88888---input "D:\Swapspace\WriteUp\Bibliography\reference.bib"
%\setlinespacing{1.44}
%\bibliographystyle{apj}
%\bibliography{../../../WriteUp/Bibliography/reference}
\bibliography{worley2009}

\end{document}